\documentclass[a4paper,10pt]{article}
\usepackage[utf8]{inputenc}
\usepackage{amssymb,amsmath}
\usepackage{authblk}
\usepackage[left=0.5in, right=0.5in, top=1in, bottom=1in]{geometry}
\usepackage{hyperref}
\usepackage{graphicx}
\title{Investigation of  Bose condensation in ideal Bose gas 
trapped under generic power law potential in 
 $d$ dimension}

\author{Mir Mehedi Faruk$^{a,b}$, Md. Sazzad Hossain$^{c}$, Md. Muktadir Rahman$^{a}$\\
Department of Theoretical Physics, University of Dhaka, Dhaka-1000$^a$\\
Theoretical Physics, Blackett Laboratory, Imperial College, London SW7 2AZ, United Kingdom$^b$\\
Department of Nuclear Engineering, University of Dhaka, Dhaka-1000$^{c}$.\\
\href{mailto:me@somewhere.com}{Email: muturza3.1416@gmail.com, mir.mehedi.faruk@cern.ch} 
 }

\begin{document}
\maketitle
 
 \begin{abstract}
The changes in characteristics of Bose condensation of ideal Bose gas due to 
an external generic power law potential $U=\sum_{i=1} ^d c_i |\frac{x_i}{a_i}|^{n_i}$
 are studied carefully. Detailed calculation of Kim $et$ $al.$
 (S. H. Kim, C. K. Kim and  K. Nahm, J Phys. Condens. Matter 11 10269 (1999).) yielded the hierarchy of
condensation transitions with changing fractional dimensionality. In this manuscript,
some theorems regarding specific heat at constant volume $C_V$ are presented. Careful examination 
of these theorems reveal the existence of hidden hierarchy of the condensation transition
in trapped systems as well.
 
 \end{abstract}

\section{Introduction}
Many authors have investigated the thermodynamic
properties of  Bose gas\cite{may,robinson,landau,huang,yan,yan1,yan2,C. J. PETHICK,ziff,beckmann,beckmann2}, particularly after it  was 
possible to create Bose Einstein Condensation (BEC) in magnetically trapped alkali-metal gases\cite{Bradley,anderson,davis}.
The constrained role of external potential does change the characteristics of quantum gases\cite{turza,sala,jellal,li,dal,toroidal}, 
providing an exciting opportunity to study the quantum mechanical effects. It is seen in the studies
that the thermodynamic behavior
of non-relativistic quantum gases are governed by polylogarithm functions both in the  case of trapped\cite{C. J. PETHICK,turza,sala,jellal} and free\cite{lee,lee1,lee2,lee3}
quantum gases.
It is also reported that the polylogarithm functions can give a single unified picture of bosons and fermions for  
free\cite{lee} and 
trapped\cite{turza3} 
  quantum gases as well. 
Both in the case of Bose and Fermi gases, different structural properties of polylogarithms\cite{yan2,sala,turza2,turza88} 
are related to different statistical effects.
The trapping potential radically changes different thermodynamic properties 
of quantum gases\cite{sala,turza,turza2,turza3}. 
Also the behavior of these quantities do change with dimensionality. 
Hence, it will be very intriguing to study the properties of Bose gas in detail before and after the 
critical temperature in arbitrary dimension with generic trapping potential of the form,
$U=\sum_{i=1} ^d c_i |\frac{x_i}{a_i}|^{n_i}$.
In this report,
we give a special emphasis on specific heat at constant volume $C_V$ as well as its derivatives, which
are the salient 
features to understand BEC\cite{kim}.\\\\
In a previous study with free Bose gas in arbitrary dimension, 
Kim $et.$ $al.$\cite{kim} reported that  
 the dimensionality contribution is a dominant factor
to specify the behavior of BEC\cite{kim}.
In general, 
there exists no discontinuity in $C_V$ at $T=T_C$ for free Bose gas in three dimensional space.
However, its derivative is  discontinuous where the magnitude of
the discontinuity is finite ($\frac{3.665 Nk}{T_C}$).
Nevertheless,
this is not true for free Bose gas in any arbitrary dimension. 
For instance, when $ d>4$
, $C_V$ is itself discontinuous at $T=T_C$\cite{pathria} for free Bose gas. In the case of trapped
Bose  gas with harmonic potential\cite{pathria,turza},
the scenario completely changes as $C_V$ becomes
 discontinuous even in $d=3$.
In order to understand 
the critical behavior of free Bose system more closely, Kim $et.$ $al.$
performed a calculation 
to check the discontinuity of $l$-th derivative of $C_V$ at $T=T_C$,
which yields the ``class" of the $C_V$ function changes with dimensionality.
This calculation shows that there exists a hierarchy of
condensation transitions with changing fractional dimensionality.
It is well known that, there is no BEC
for free Bose gas in $d\leq2$ and   Kim $et.$ $al.$ 
found $C_V$ to be smooth function in the whole temperature range in this situation. 
If  the dimensionality ranges from $2<d\leq3$, the discontinuity of $l$-th derivative of $C_V$ 
at $T=T_C$ depends upon the sub-interval we choose in this range (see Table 1 of Ref. \cite{kim}),
while the first derivative of $C_V$ remains discontinuous for $d>3$.   
In this report, we have proposed a new theorem, concerning the $l$-th derivative of $C_V$
for  Bose gas trapped under generic power law potential,
which coincides with the calculation of Kim $et.$ $al.$, in the limit all $n_i\rightarrow\infty$.
Therefore one can can reproduce the  previous calculation as
a special case of the new theorem. This new theorem helps us to
understand the hidden hierarchy of condensation transition,
existing in trapped systems as well. Most importantly, one can find situations 
even in an integer dimension
where the $l$-th derivative of 
$C_V$ of trapped Bose gas is
continuous, with appropriate trapping potential.
\\\\
In this work,  the grand potential is determined at first,
from which all the  thermodynamic quantities such as internal energy $E$, entropy $S$, Helmholtz free energy $F$ and
$C_V$ are calculated. In order to scrutinize them closely, the fugacity (chemical potential) is evaluated numerically
in arbitrary dimension
with any trapping potential
as a function of reduced temperature $\tau=\frac{T}{T_C}$ (see appendix) following Kelly's work\cite{kelly}.
Then we propose the theorems regrading $C_V$
and its derivative.
Beside this, the latent heat of condensation transition is studied in detail from the Clausius-Clapayron equation\cite{huang}, 
from which we have deduced the required 
latent heat
for Bose condensation in arbitrary dimension with any trapping potential.
All the calculated quantities 
reduce to text book result of free Bose gas when all $n_i\rightarrow \infty$\cite{pathria}.\\\\
The report is organized in the following way. The grand potential and the other thermodynamic quantities are
presented
in section 2 and 3, respectively. In section 4 we deduce $C_V$ and the significant theorems regarding this. Section 5 is
devoted to investigate the latent heat of condensation for trapped system. 
Results are discussed  in section 6. The report is
concluded in section 7.

\section{Density of States and grand potential}
Considering
the ideal  Bose gas in a confining external potential in a d-dimensional space with Hamiltonian, 
\begin{eqnarray}
\epsilon (p,x_i)= bp^l + \sum_{i=1} ^d c_i |\frac{x_i}{a_i}|^{n_i}
\end{eqnarray}
Where,  $p$ is the momentum 
and $x_i$ is the  $i$ th component of coordinate of a particle
and
$b,$ $l,$ $a_i$, $c_i$, $n_i$  are all positive constants. Here, $c_i$, $a_i$ and $n_i$ determine the depth 
and confinement power of
the potential. Using $l=2$, $b=\frac{1}{2m}$ one can get the energy spectrum  of  the Hamiltonian used in 
the literature \cite{pathria,huang,ziff,sala,yan2}. Note that, $x_i<a_i$.
For the free system all $n_i\longrightarrow \infty$. As $|\frac{x_i}{a_i}|<1$,
the potential term goes to zero when all $n_i\longrightarrow \infty$.  \\\\
The resulting density of states with this Hamiltonian is\cite{yan2},
\begin{eqnarray}
\rho(E)=C(m,V_d')E^{\chi-1} 
\end{eqnarray}
where, $C(m,V_d')$ is a constant depending on effective volume $V_d'$ and mass $m$\cite{yan2,turza} \\\\
The grand potential for the Bose system, 
\begin{eqnarray}
q=-\sum_\epsilon ln(1-zexp(-\beta \epsilon)) 
\end{eqnarray}
where $k$, $\mu$ and  $z=\exp(\beta \mu)$ being the Boltzmann 
Constant, the chemical potential and fugacity respectively and
$\beta=\frac{1}{kT}$.\\\\
Replacing the sum by integration we get,
\begin{eqnarray}
 q=q_0 -\int_0 ^\infty \rho(\epsilon) ln(1-zexp(-\beta \epsilon))
\end{eqnarray}
which yields,\cite{yan2,turza}
\begin{equation}
 q=q_0+g\frac{V_d'}{\lambda '^d}g_{\chi+1}(z)
\end{equation}\\
Where, \begin{eqnarray}
 g_l(z)&=&\int_0 ^\infty dx\frac{x^{l-1}}{z^{-1}e^x-1}=\sum_{j=1} ^\infty\frac{z^j}{j^l}\\
                q_0&=&-ln(1-z)\\   
                V_d ' &=& V_d \prod_{i=1} ^d (\frac{kT}{c_i})^{1/n_i}\Gamma(\frac{1}{n_i} + 1)\\
                \lambda'&=& \frac{h b^{\frac{1}{l} }}{\pi ^{\frac{1}{2}} (kT) ^{\frac{1}{l}}} [\frac{d/2+1}{d/l+1}]^{1/d}\\
                \chi&=& \frac{d}{l} + \sum_{i=1} ^d \frac{1}{n_i}
               \end{eqnarray}
And, $ V_d '$ and $g_{\chi}(z)$
is known as effective volume and Bose function respectively.
Now, as $z\rightarrow 1$, the Bose function $g_\chi(z)$ approaches $\zeta(\chi)$, for $\chi>1$\cite{pathria}.
Another representation of Bose
function $g_\chi(z)$ due to Robinson is \cite{pathria,kim}
\begin{eqnarray}
 g_\chi(e^{-\eta})=\frac{\Gamma(1-\chi)}{\eta ^{1-\chi}} + \sum_{i=0} ^{\infty}\zeta(\chi-i)\eta^i
\end{eqnarray}

\section{Thermodynamic quantities}
The number of particles can be evaluated from the grand potential\cite{yan2,turza},
\begin{eqnarray}
 N&=&z(\frac{\partial q}{\partial z})_{\beta,V} \nonumber\\
\Rightarrow N-N_0&=& g \frac{V_d '}{{\lambda '} ^d}g_\chi(z)
\end{eqnarray}
The above equation suggests the only relevant quantity  that determines BEC to
take place is $\chi$. In case of BEC as $T\rightarrow T_C$, $z\rightarrow 1$.
So, BEC would take place when,
\begin{eqnarray}
\chi=\frac{d}{l}+\sum_i ^d \frac{1}{n_i}>1 
\end{eqnarray}
And the critical temperature is,
\begin{equation}
 T_c=\frac{1}{k}[\frac{N_c h^d b^{d/l}  \prod_{i=1} ^d c_i^{1/n_i} }{gC_n\Gamma(\frac{d}{l} +1)
 V_d \prod_{i=1} ^d  \Gamma(\frac{1}{n_i}+1 ) \zeta(\chi)}]^{\frac{1}{\chi} }
\end{equation}\\
For massive bosons (with $l=2$ ),
one can find the BEC criterion,
\begin{equation}
 \frac{d}{2}+\sum_i ^d \frac{1}{n_i}>1
\end{equation}
Therefore, for free massive bosons ($\sum_i ^d \frac{1}{n_i}\rightarrow 0$), the above criterion reduces to
\begin{equation}
 \frac{d}{2}>1
\end{equation}
which coincides with the literature\cite{pathria}.\\\\
The  ground state fraction \cite{yan2,turza}
from eq. (12) and (14),
\begin{eqnarray}
 \frac{N_0}{N}=1-(\frac{T}{T_c})^\chi
\end{eqnarray}\\
\begin{figure}[h!]
\centering
\includegraphics[ height=8cm, width=8cm]{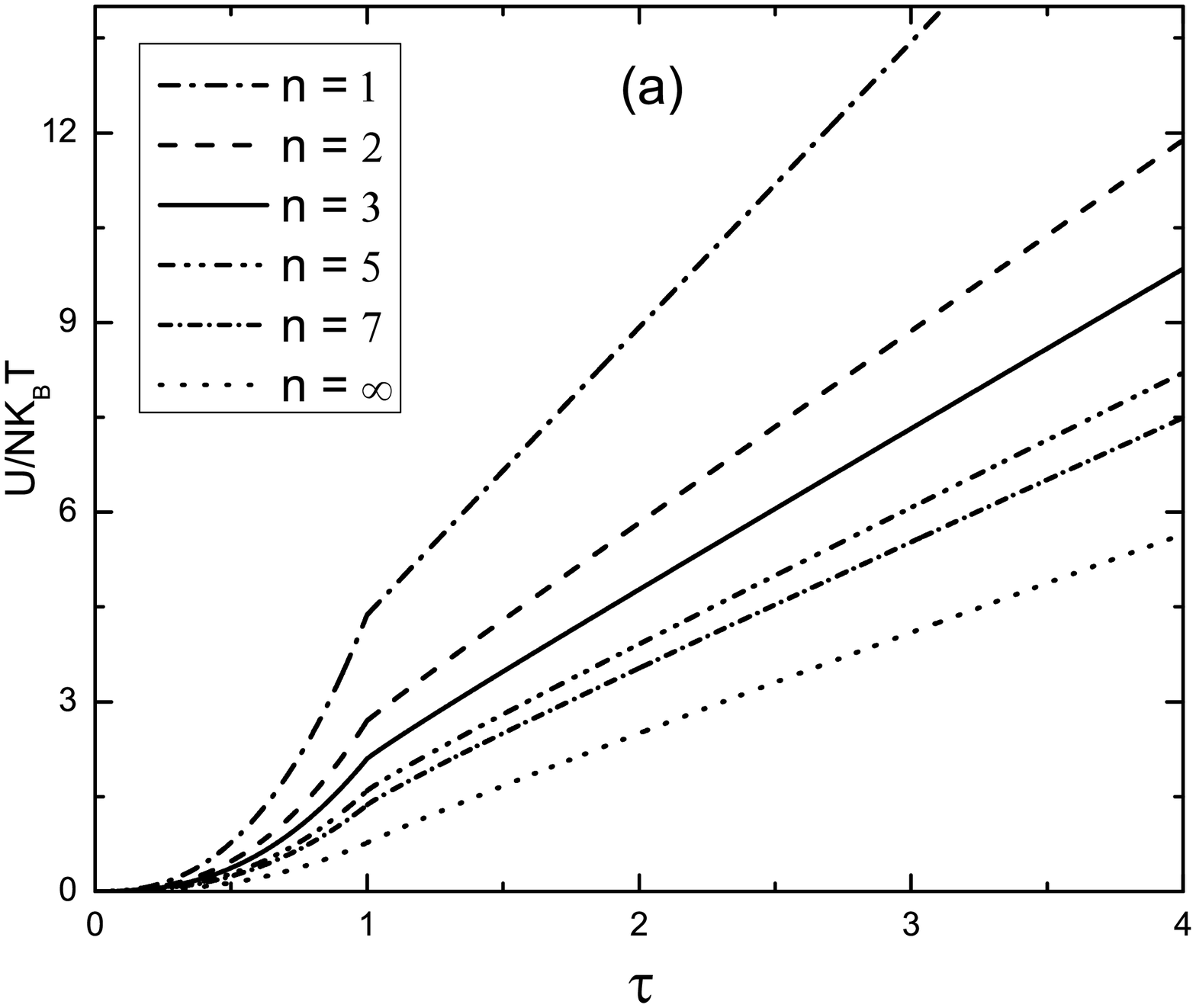}  
\includegraphics[ height=8cm, width=8cm]{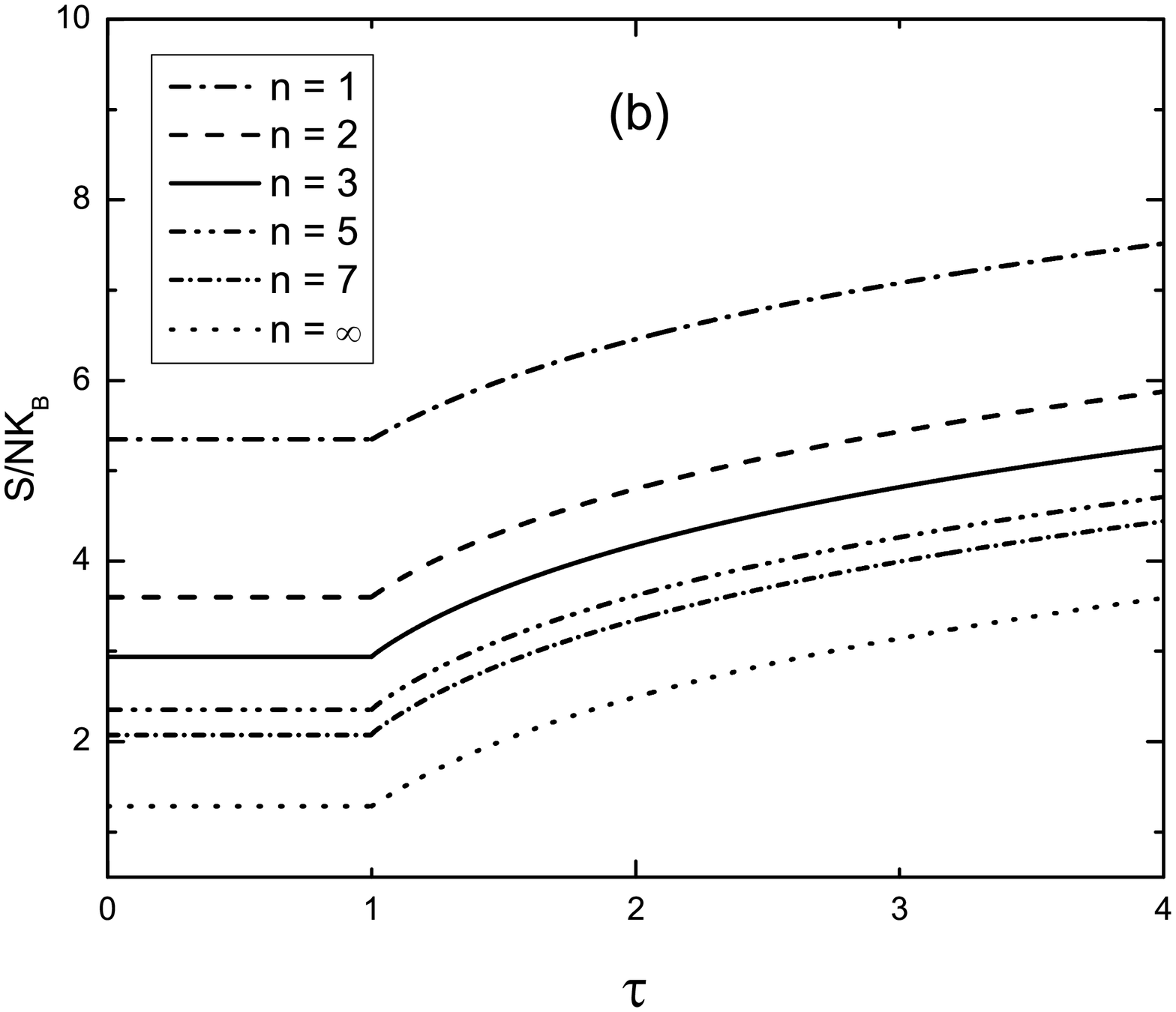}  
\includegraphics[ height=8cm, width=8cm]{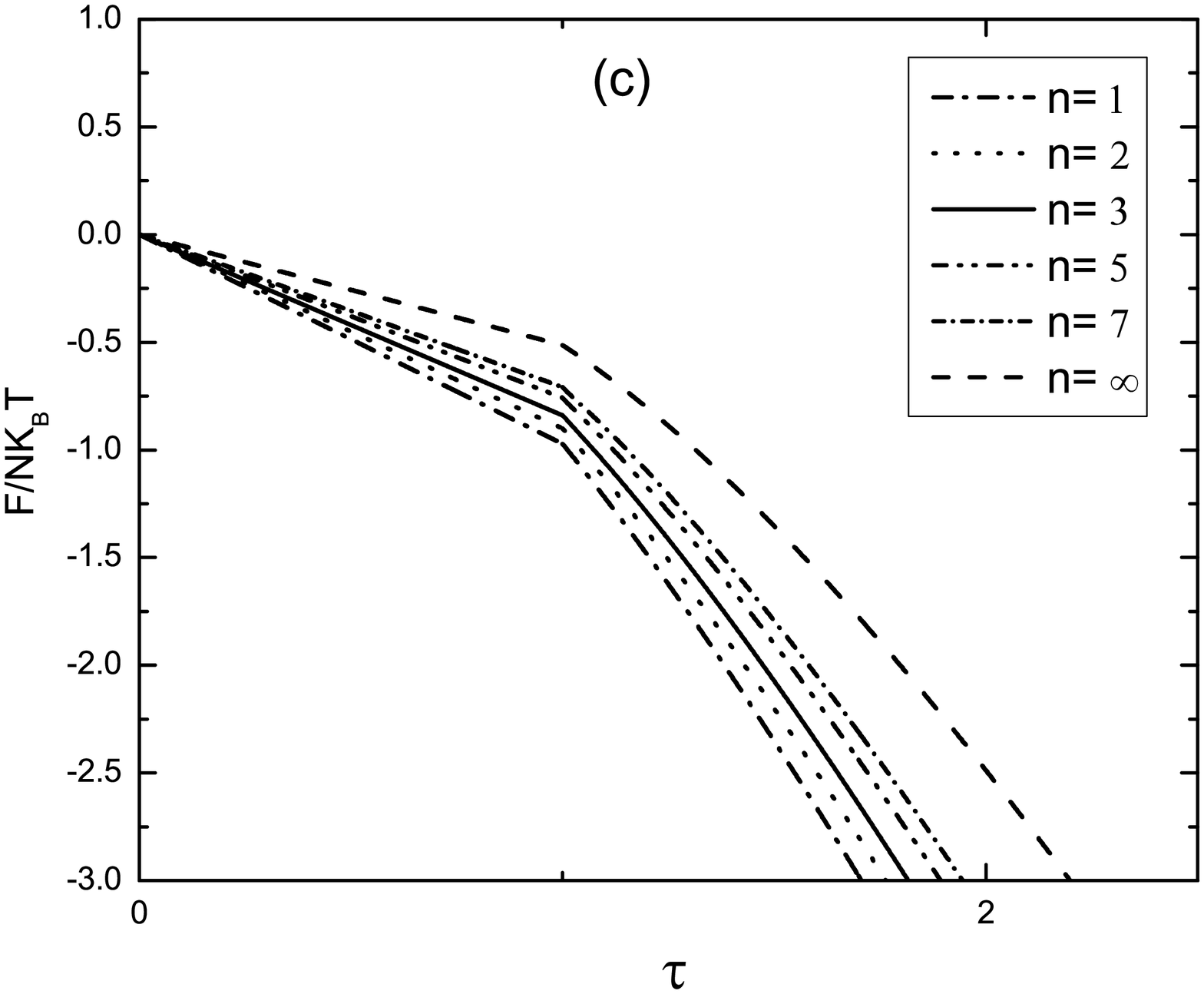}
\includegraphics[ height=8cm, width=8cm]{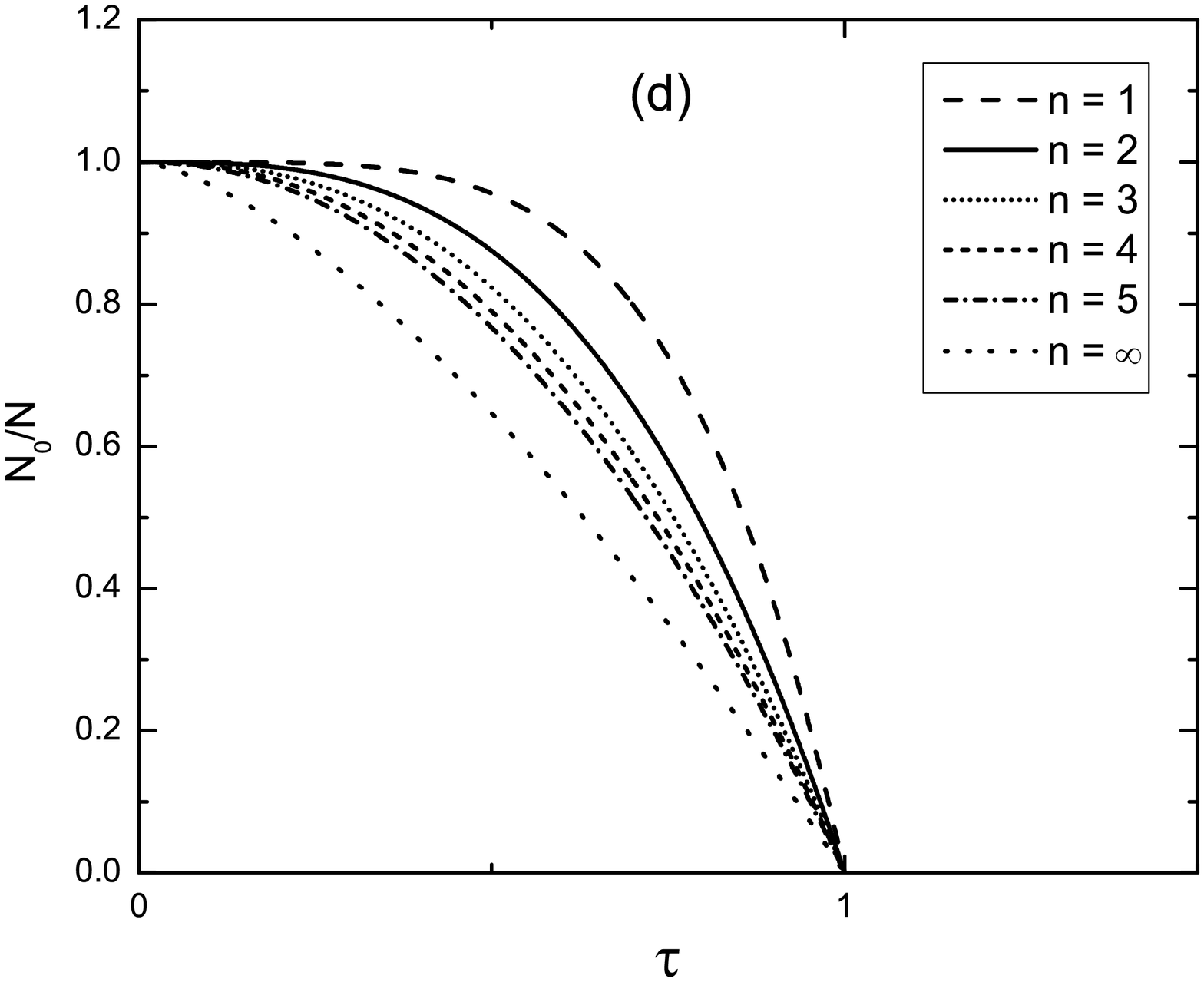}
\caption{Fig (a), (b), (c) and (d) respectively represents internal energy, entropy, 
Helmholtz free energy and ground state fraction of Bose system, with different trapping potential in $d=3$
. $n=\infty$ denotes free Bose system.}
  \label{fig:boat1}
\end{figure}
The above equation produces the exact 
ground state fraction  for free system 
when all $n_i\rightarrow \infty$\cite{pathria}.
Again, obtaining internal energy $E$ \cite{yan2,turza}
from the grand potential, 
\begin{eqnarray}
   E =-(\frac{\partial q}{\partial \beta})_{z,V}= \left\{
     \begin{array}{lr}
       NkT \chi \frac{g_{\chi+1}(z)}{g_{\chi}(z)} &,  T>T_c\\
       N kT \chi \frac{\zeta(\chi+1)}{\zeta(\chi)} (\frac{T}{T_c})^\chi&,  T \leq T_c
     \end{array}
   \right.
\end{eqnarray}                       \\
Now, the entropy $S$\cite{yan2,turza}, below and greater than the critical temperature
     \begin{eqnarray}
   S=kT(\frac{\partial q}{\partial T})_{z,V} -Nk\ln z +kq = \left\{
     \begin{array}{lr}
       N k [\frac{v_d'}{\lambda '^d}(\chi+1) {g_{\chi + 1}(z)}-\ln z] &,  T>T_c\\
       (N-N_0) k \frac{v_d'}{\lambda '^d}(\chi+1) \zeta(\chi+1)&,  T\leq T_c
     \end{array}
   \right.
\end{eqnarray}\\
And, finally the expression of Helmholtz free energy,
\begin{eqnarray}
    A &&= -kTq+NkT\ln z\nonumber\\
   \frac{A}{NkT} &&= \left\{
     \begin{array}{lr}
      -  \frac{g_{\chi+1} (z)}{g_{\chi} (z)} + \ln z&,  T>T_c\\
      -  \frac{\nu}{\lambda ^d}\zeta(\chi+1) &,  T\leq T_c
     \end{array}
   \right.
\end{eqnarray}         \\
\section{Heat  capacity at constant volume $C_V$}
Heat capacity at constant volume $C_v$ below and above $T_c$ \cite{yan2,turza},
\begin{eqnarray}
   C_V =T(\frac{\partial S}{\partial T})_{N, V} =\left\{
     \begin{array}{lr}
       N k [\chi(\chi+1)\frac{g_{\chi+1}(z)}{g_{\chi}(z)}-\chi^2 \frac{g_{\chi}(z)}{g_{\chi-1}(z)}] &,  T>T_c\\
       N k \chi (\chi +1)\frac{\zeta(\chi+1)}{\zeta(\chi)} (\frac{T}{T_C})^\chi&,  T\leq T_c
     \end{array}
   \right.
\end{eqnarray}  \\\\
Another important quantity $\frac{\partial C_{v}}{\partial T}$
below and above $T_C$,
\begin{eqnarray}
  \frac{1}{N k } \frac{\partial C_{v}}{\partial T}  =\left\{
     \begin{array}{lr}
        \frac{1}{T}\Big[\chi^{2}(\chi+1)\frac{g_{\chi+1}(z)}{g_{\chi}(z)}
-\chi^{2}\frac{g_{\chi}(z)}{g_{\chi-1}(z)}
-\chi^{3}\frac{g_{\chi}(z)^{2}g_{\chi-2}(z)}{g_{\chi-1}(z)^{3}}\Big] &,  T>T_c\\
\frac{1}{T}\chi^2(\chi+1)\frac{\zeta(\chi+1)}{\zeta(\chi)}(\frac{T}{T_C})^{\chi}        &,  T\leq T_c
     \end{array}
   \right.
\end{eqnarray}  \\\\
\begin{figure}[h!]
\centering  
\includegraphics[ height=8cm, width=8cm]{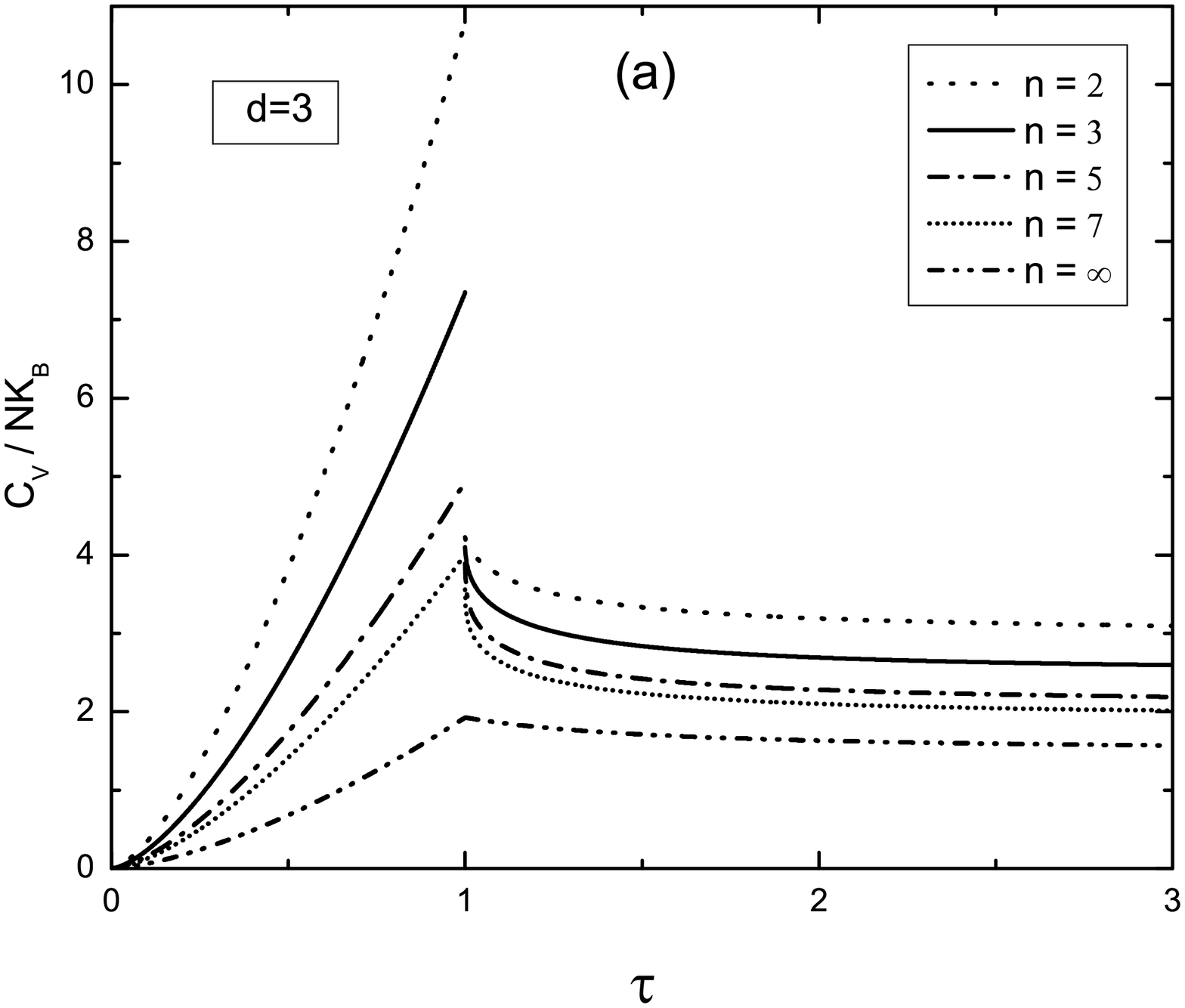}  
\includegraphics[ height=8cm, width=8cm]{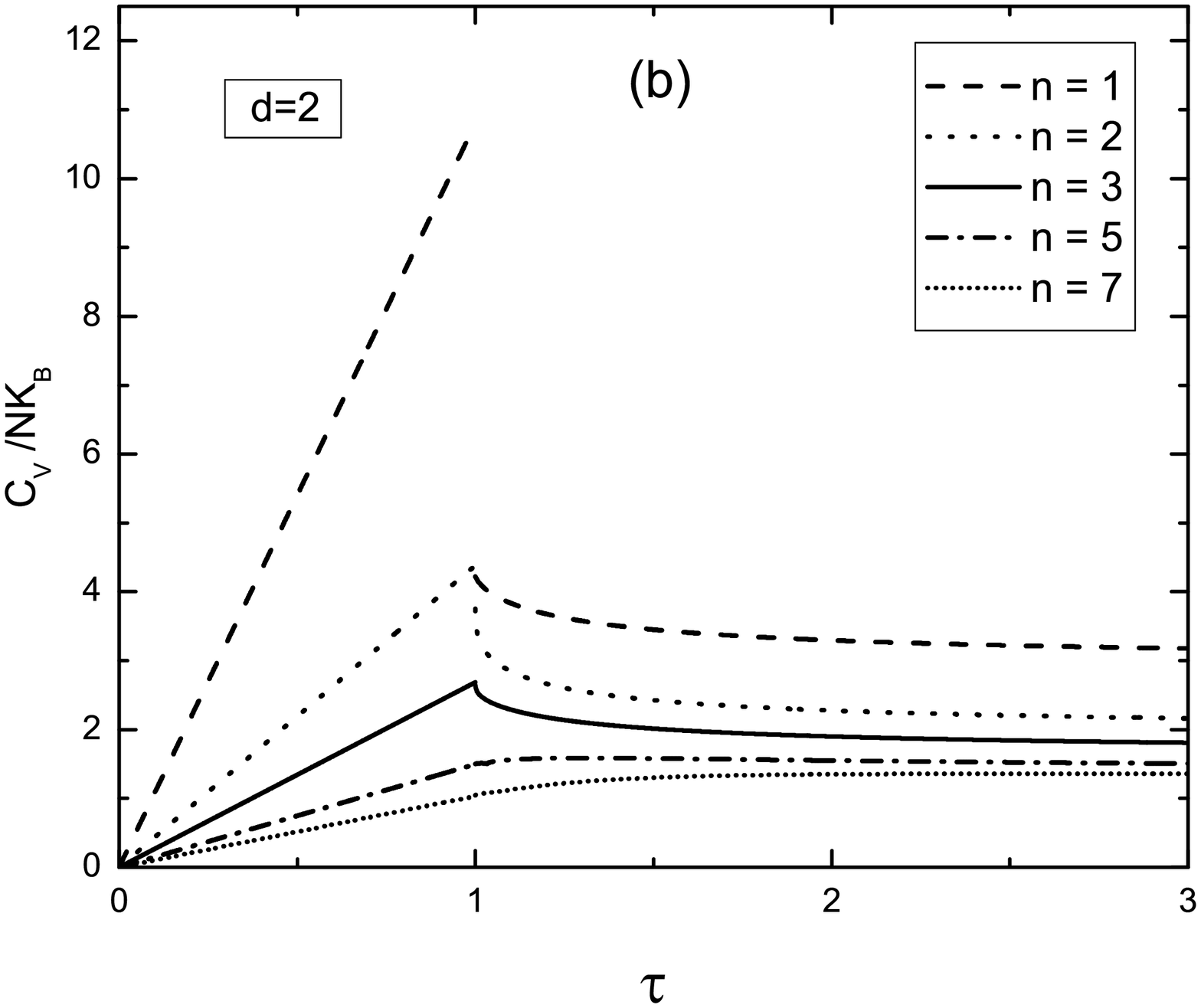}
\caption{Fig (a) and (b),   represents $C_V$ of
Bose system, with different trapping potential in $d=3$ and $d=2$, respectively.
}
  \label{fig:boat1}
\end{figure}\\\\ 
\textbf{Theorem 4.1}: {\em Let an ideal Bose gas in an external potential, $U=\sum_{i=1} ^d c_i |\frac{x_i}{a_i}|^{n_i}$, \\\\
(i) if the Bose gas does condense ($\chi>1$), heat capacity
$C_V$ will be discontinuous at $T=T_c$
if and only if,
\begin{eqnarray}
\chi=\frac{d}{l}+\sum_{i=1} ^d \frac{1}{n_i}>2 \nonumber
\end{eqnarray}}. \\
(ii) And the difference between the heat capacities at constant volume, at $T_c$ as
\begin{eqnarray}
\Delta C_V\mid_{_{T=T_c}}=C_V\mid_{_{T_c^{-}}}- C_V\mid_{_{T_c^{+}}}=Nk\chi^2\frac{\zeta(\chi)}{\zeta(\chi-1)}\nonumber
\end{eqnarray}
\\\\
\textbf{Proof}:\\\\
As $T\rightarrow T_c$, $z\rightarrow 1$ and $\eta\rightarrow 0$, where $\eta=-\ln z$.
For $T\rightarrow T_c ^+$, 
\begin{eqnarray}
C_V (T_C^+) &&=     N k [\chi(\chi+1)\frac{\nu'}{\lambda '^D}g_{\chi+1}(z)|_{_{z\rightarrow 1}}
-\chi^2 \frac{g_{\chi}(z)}{g_{\chi-1}(z)}|_{_{z\rightarrow 1}}] \nonumber \\
&& =N k [\chi(\chi+1)\frac{\nu'}{\lambda '^D}\zeta({\chi+1})- \chi^2 \frac{\zeta({\chi})}{g_{\chi-1}(z)}|_{_{z\rightarrow 1}}]\
\end{eqnarray}
As the denominator of the second term of the right hand side 
contains $g_{\chi-1}(z)$, we can not simply substitute it with zeta function as $z\rightarrow 1$
. So, using the  representation of Bose function by Robinson (Eq. 11),
we get from the above
\begin{eqnarray}
 C_V (T_C^+) = N k [\chi(\chi+1)\frac{\nu'}{\lambda '^D}\zeta({\chi+1})-
 \chi^2 \frac{\zeta({\chi})}{\Gamma(2-\chi)} \eta ^{2-\chi}\mid _{\eta\rightarrow 0}]
\end{eqnarray}
On the other hand 
\begin{eqnarray}
 C_V (T_C^-) = N k [\chi(\chi+1)\frac{\nu'}{\lambda '^D}\zeta({\chi+1})
\end{eqnarray}\\
Taking the difference between $C_V(T_C ^+)$ and $C_V(T_C ^-)$, we get,
\begin{eqnarray}
\Delta C_V\mid_{_{T=T_c}}=  \chi^2 \frac{\zeta({\chi})}{\Gamma(2-\chi)} \eta ^{2-\chi}\mid _{\eta\rightarrow 0}
\end{eqnarray} 
Which dictates, $C\mid_{_{T=T_c}}$ 
will be non zero for $\chi>2$. So, $C_V$ will be discontinuous
when $\chi>2$
and thus completing first part of the theorem.\\\\
As, 
$\chi$ should be greater than two for $\Delta C_V$ at $T=T_C$ to be non-zero, we can re-write 
equation (23), by substituting $g_{\chi-1}(z)$ by zeta function.
\begin{eqnarray}
C_V (T_C^+) = N k [\chi(\chi+1)\frac{\nu'}{\lambda '^D}\zeta({\chi+1})- \chi^2 \frac{\zeta({\chi})}{\zeta({\chi-1})}]
\end{eqnarray}
Now, from Eq. (25) and (27) we can write.
\begin{eqnarray}
\Delta C_V\mid_{_{T=T_c}}=C_V\mid_{_{T_c^{-}}}- C_V\mid_{_{T_c^{+}}}=Nk\chi^2\frac{\zeta(\chi)}{\zeta(\chi-1)}\nonumber
\end{eqnarray}\\
In case of free Bose gas the first part of the theorem dictates there will be a jump in $C_V$ at
$T=T_C$ for $d>4$, producing   the same result as  Ziff $et$ $al.$       \\\\
\textbf{Theorem 4.2}: {\em Let an ideal Bose gas in an external potential, $U=\sum_{i=1} ^d c_i |\frac{x_i}{a_i}|^{n_i}$, \\\\
(i) the jump of the the first derivative of $C_V$ at $T=T_C$ will be finite for $\chi=\frac{3}{2}$,
it will be infinite for $\chi>\frac{3}{2}$ and no jump for $\chi<\frac{3}{2}$\\\\
(ii) And the finite jump of the first derivative of $C_V$, at $T=T_c$ is
\begin{eqnarray}
\Delta \frac{\partial C_V}{\partial T}\mid_{_{T=T_c}}=\frac{\partial C_V}{\partial T}\mid_{_{T_c^{-}}} -
\frac{\partial C_V}{\partial T}\mid_{_{T_c^{+}}}=\frac{27Nk}{8T_{c}}
[\zeta(\frac{3}{2})]^2\frac{\Gamma(\frac{3}{2})}{\Gamma(\frac{1}{2})}
\nonumber
\end{eqnarray}}\\\\
\textbf{Proof}:\\\\
From equation (21) as $T\longrightarrow T_{c}$, we obtain,
\begin{eqnarray}
  \frac{1}{N k } \frac{\partial C_{v} }{\partial T}  =\left\{
     \begin{array}{lr}
        \frac{1}{T_{c}}\Big[\chi^{2}(\chi+1)\frac{g_{\chi+1}(z)}{g_{\chi}(z)}
-\chi^{2}\frac{g_{\chi}(z)}{g_{\chi-1}(z)}
-\chi^{3}\frac{g_{\chi}(z)^{2}g_{\chi-2}(z)}{g_{\chi-1}(z)^{3}}\Big] &,  T\rightarrow T_{c}^{+}\\
\frac{1}{T_{c}}\chi^2(\chi+1)\frac{\zeta(\chi+1)}{\zeta(\chi)}        &,  T\rightarrow T_c^{-}
     \end{array}
   \right.
\end{eqnarray}
Now taking the difference,
\begin{eqnarray}
\Delta \frac{\partial C_V}{\partial T}\mid_{_{T=T_c}}=\frac{Nk}{T_{c}}\Big[\chi^{2}\frac{g_{\chi}(z)}{g_{\chi-1}(z)}
+\chi^{3}\frac{g_{\chi}(z)^{2}g_{\chi-2}(z)}{g_{\chi-1}(z)^{3}}\Big]\mid_{_{z\rightarrow 1}}
\end{eqnarray}
Using the representation due to Robinson,
\begin{eqnarray}
\Delta \frac{\partial C_V}{\partial T}\mid_{_{T=T_c}}=\frac{Nk}{T_{c}}\Big [\chi^{2}\frac{ \zeta(\chi)
}{\Gamma(2-\chi)}\eta^{2-\chi}
+\chi^{3}\frac{\Gamma(3-\chi)}{[\Gamma(2-\chi)]^{3}}[\zeta(\chi)]^{2}\eta^{3-2\chi}]\mid_{\eta\rightarrow0}
\end{eqnarray}
\\
Therefore the above equation suggests, $\Delta \frac{\partial C_V}{\partial T}\mid_{_{T=T_c}}$
will be nonzero when $\chi\geq 3/2$ and zero elsewhere.
Now, as $\chi>\frac{3}{2}$ the second term obviously diverges towards infinity and making the whole term infinite.
But as $\chi=\frac{3}{2}$, the first term vanishes as $\eta\rightarrow 0$ and the second term becomes finite. 
Hence, completing the first part of the proof.
Putting $\chi=\frac{3}{2}$ in Eq. (30) yields
\begin{eqnarray}
\Delta \frac{\partial C_V}{\partial T}\mid_{_{T=T_c}}
= (\frac{3}{2})^{3}\frac{\Gamma(3-\frac{3}{2})}{[\Gamma(2-\frac{3}{2})]^{3}}[\zeta(\frac{3}{2})]^{2}]
=\frac{27Nk}{8T_{c}}
[\zeta(\frac{3}{2})]^2\frac{\Gamma(\frac{3}{2})}{\Gamma(\frac{1}{2})}
\end{eqnarray}
Which completes the second part of the proof. 
In case of free massive (all $n_i\rightarrow \infty$) bosons in $d=3$ Eq. (31) yields,
the magnitude  of the discontinuity of first derivative of $C_V$
at $T=T_C$ is $3.665\frac{Nk}{T_C}$, reproducing the exact same result of Pathria\cite{pathria}.
\\\\
\textbf{Theorem 4.3}: {\em Let an ideal Bose gas in an external potential, $U=\sum_{i=1} ^d c_i |\frac{x_i}{a_i}|^{n_i}$, 
The between $l$-th derivative of heat capacities at $T=T_C$ is,
\begin{eqnarray}
 \Delta^{l}(T_C)= \lim_{T\rightarrow T_C}[(\frac{\partial}{\partial T})^{l}C_{v}^{-}(T)-(\frac{\partial}{\partial T})^{l}C_{v}^{+}(T)]
=\lim_{\eta \longrightarrow 0} \sum_{j=1} ^{l} a_{lj}\eta^{j+2-(j+1)\chi} \nonumber
\end{eqnarray}}\\
\textbf{Proof:}\\ \\
We prove the above equation by the method of induction. \\\\
 For $l=1$,
\begin{eqnarray}
 \Delta^{1}(T_{c})=\lim_{\eta\rightarrow 0}a_{11}\eta^{3-2\chi}
\end{eqnarray}
This is clearly true from equation (13) and the known result for $D = 3$.\\ \\
 Let's assume the equation holds for any positive integer $l$. Then,
\begin{eqnarray}
 \Delta^{l}(T_{c})=\lim_{\eta\rightarrow 0}\sum_{j=1} ^{l}a_{lj}\eta^{j+2-(j+1)\chi}
\end{eqnarray}
Now, in case of $l+1$,
\begin{eqnarray}
 \Delta^{l+1}(T_c)&=& \lim_{\eta\rightarrow 0}\sum_{j=1} ^{l} a_{lj}(j+2-(j+1)\chi)\frac{\chi\zeta(\chi)}{T_{c}
 \Gamma (2-\chi)} \eta^{j+3-(j+2)\chi} \nonumber\\
&=& \lim_{\eta\rightarrow 0}\sum_{j=2}^{k+1} a_{l+1, j-1}(j+1-j\chi)\frac{\chi\zeta (\chi)}
{T_{c}\Gamma(2-\chi)}\eta^{j+2-(j+1)\chi} \nonumber\\
&=& \lim_{\eta\rightarrow 0}\sum_{j=2}^{l+1} a_{l+1,j} \eta^{j+2-(j+1)\chi}
 \end{eqnarray}
$a_{lj}$ satisfies the recurrence relation.
\begin{eqnarray}
 a_{l+1,j}= \frac{({j+1-j\chi})\chi \zeta(\chi)}
 {T_{c} \Gamma ({{2-\chi}})} a_{l+1,j-1} \\
\end{eqnarray}
where, $j=2,3,4,...,k+1$.\\\\
Therefore, (a) and (b) enable us, for any positive integer $l$, to write
\begin{eqnarray}
 \Delta ^{l}(T_{c})=\lim_{\eta \longrightarrow 0} \sum_{j=1} ^{l} a_{lj}\eta^{j+2-(j+1)\chi}
\end{eqnarray}
This above equation coincides with Kim $et.$ $al.$ in case of free massive boson. 
\\\\
\section{Latent Heat}
\begin{figure}[h!]
\centering
\includegraphics[ height=8cm, width=8cm]{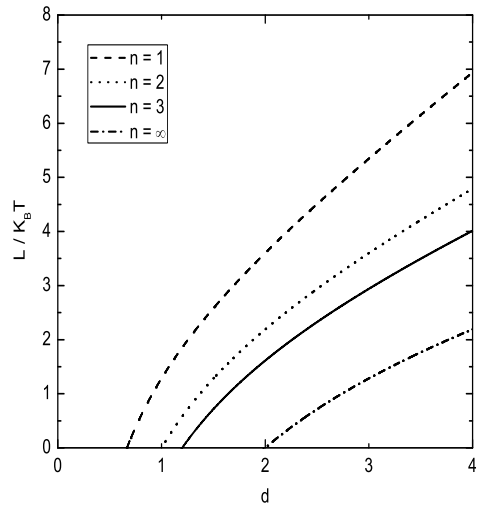}  
 \caption{Latent heat of Bose condensation }
  \label{fig:boat1}
\end{figure}
Just as,
any first order phase transition pressure is governed by Clausius-Clapeyron equation, 
in the transition line\cite{huang,ziff}. 
Like the BEC for free Bose gas at $d = 3$, BEC for trapped Bose gas do also exhibit
first order phase transition as they obey the Clausius-Clapayron equation\cite{huang}.
The Clausius-Clapeyron equation which is derived from Maxwell relations take the form,
\begin{eqnarray}
 \frac{dP}{dT}=\frac{\Delta s}{\Delta v}=\frac{L}{T\Delta v}
\end{eqnarray}
where, $L$, $\Delta s$ and $\Delta v$ are the latent heat, change in entropy and change in volume respectively. 
The effective  pressure in phase transition line is,
\begin{equation}
 P_0(T)=\frac{kT}{\lambda'^d}g_{\chi +1}(1)
\end{equation}
Differentiating with respect to $T$ leads, 
\begin{equation}
 \frac{dP_0}{dT}=\frac{1}{Tv_g}[(\chi+1)kT\frac{g_{\chi+1}(1)}{g_{\chi}(1)}]
\end{equation}\\
When two phases coexist
the non condensed  phase has specific volume $v_g$ whether the condensed phase 
has specific volume has specific 
volume  $0$,  concluding $\Delta v=v_g$.
So, the latent heat of transition per particle in case of
trapped Boson is
\begin{eqnarray}
 L=(\chi+1)\frac{\zeta{(\chi +1)}}{\zeta({\chi })}kT
\end{eqnarray}\\
So, 
in the 
case of free massive boson the latent heat per particle in three dimensional space is,
\begin{equation}
 L=\frac{5}{2}kT\frac{\zeta(5/2)}{\zeta({3/2 })}
\end{equation}
same as the text book results\cite{huang}.

\section{Discussion}
In this section we discuss the influence of trapping potential on thermodynamic quantities.
Beside this we also explain the theorems and their implications as well. In the figures, we have used $n_1=n_2=..=n_i=..=n_d=n$ (isotropic potential),
but the formulas described in the above sections  are more general.
\\\\
In Fig (1), we have described the influence of different power law potentials on
thermodynamic quantities such as internal energy $E$, entropy $S$, free energy $F$ and ground state fraction
of Bose gas in three dimensional space. 
 In the case of internal energy, a strict nonlinear behavior is 
observed when $T<T_C$ and this behavior is more noticeable when we decrease the value of $n$. But 
in principle, the effect of power law potential is seen in both below and above $T_C$. 
Same phenomena is also observed in the case of $S$ and $F$. 
Entropy remains constant in the condensed phase, with a specific choice of $n$. 
But the entropy increases 
while the value of free energy gets lower with trapping potential. Now turning our attention towards ground state fraction, 
Eq. (15) and (17) dictates, $|\frac{dN_0}{dT}|_{T=T_c}|>\frac{N}{T_C}$. Thus, this relation depicts a very significant characteristic at the onset of BEC,
that the $N_0-T$ curves are always convex for Bose system in condensed phase.
But their curvatures are different depending on trapping potentials as shown in figure 1 (d). \\\\
Fig 2 (a), (b) illustrates the $C_V$ of Bose system 
with different power law potentials in three and two dimensional space, respectively. 
When $d=3$, there is no discontinuity  in free system for $C_V$ at $T=T_C$. But according to theorem 4.1 in three dimensional space, $C_V$ becomes 
discontinuous when  $n<6$, for isotropic trapping  potential, which is visible in fig. 2 (a).
In the case of $d=2$, theorem A dictates that $C_V$ becomes discontinuous 
when $n<2$ for isotropic potential, also conspicuous in figure 2 (b). Therefore, we can conclude that theorem A can exactly govern the discontinuity condition of $C_V$ at $T=T_C$.
Now, let us turn our attention towards 
latent heat. Fig 3 illustrates the behavior of latent heat of trapped Bose gas, with changing dimensionality.
As BEC is a 1st order phase transition\cite{huang}, latent heat is associated with this  phase transition.
So, zero latent heat denotes no phase transition  $i.e.$ no condensed phase.
As there is no phase transition in $d<2$ for free Bose gas, 
there should be no latent heat associated with $d<2$ in the case of free system,
which is manifested in figure 3.
But the scenario changes when we apply trapping potential.
From Eq. (13) we can say,
there will be condensed phase in $d=2$ for any trapping potential unless $\sum_{i=1,2}\frac{1}{n_i}=0$.
Fig. 3 is  in accordance with this fact as latent
heat is seen to be
non-zero with any trapping potential.
In the case of $d=1$ we get from Eq. (13)
that BEC will take place if $n<2$. It is seen in the Fig.
3 that latent heat is non-zero  when $n=1$ indicating the existence of condensed phase in one dimensional space.
\\
\begin{table}[h!]

\caption{The hierarchy of the condensation transition of ideal Bose gas trapped under generic power law potential.
The result of table are in agreement with Kim $et$ $al.$ for free system. But in the case of trapped system the ''class"
of functions significantly change depending on the values of $n_i$}
\begin{center}

\begin{tabular}
{ c || c c c c c c c }
\hline 
$\chi$  & $(\partial / \partial T)C_{v} $ & $(\partial / 
\partial T)^{2}C_{v} $ & $(\partial / \partial T)^{3}C_{v} $ & $(\partial / \partial T)^{4}C_{v} $ & ..... & Class\\ \hline 
$\chi=\frac{3}{2}$  & d  & & & & & $C^{0}$ \\
$\frac{4}{3}\leq \chi<\frac{3}{2}$ & c  & d & & & & $C^{1}$ \\
$\frac{5}{4} \leq \chi<\frac{4}{3}$  & c  & c & d & & & $C^{2}$ \\
. & & & & & & & \\
. & & & & & & & \\
. & & & & & & & \\
$\frac{l+2}{l+1} \leq \chi<\frac{l+1}{l}$  & c  & c & c & & ...(d) & $C^{l-1}$ \\
. & & & & & & & \\
. & & & & & & & \\
. & & & & & & & \\
$\chi=1$  & c  & c & c & c & c & $C^{\infty}$ \\
\hline  
\end{tabular}

\end{center}

\end{table}

Let us now concentrate on the significance of theorem 4.3. The calculation done by Kim 
$et.$ $al.$ shows how the condensed phases can be different, depending on the dimensionality for free Bose gas. 
The dimensional dependence of  discontinuity of the $l$ th derivative of $C_V$  
indicates hidden hierarchy of the condensation transition with changing fractional dimensionality. Theorem 4.3 generalizes this result for trapped systems indicating a 
similar hierarchy where we find the class of $l^{th}$ derivative of $C_{v}$ depends on
$\chi$. We now elaborate how this theorem  classifies the class of $l^{th}$ derivative of $C_{v}$ and present it in table 1.\\\\
(i) From Eq. (37), when
 $l=1$, we see the difference between 1st derivative before and after $T_{c}$,
\begin{eqnarray}
\Delta=a_{11}\eta^{3-2\chi}\lvert_{\eta\rightarrow 0} \nonumber
\end{eqnarray}
In order to be discontinuous we need $\Delta \neq 0$,
which will be true, when
 $3-2\chi \leq 0$ i.e. $\chi \geq \frac{3}{2}$.
Furthermore, if $\chi > \frac{3}{2}$, the 1st derivative is infinitely discontinuous and  $\chi=\frac{3}{2}$ denotes 
 finite discontinuity.\\ \\
(ii) Again, for $l=2$,
\begin{eqnarray}
\Delta=a_{21}\eta^{3-2\chi}+a_{22}\eta^{4-3\chi}\lvert_{\eta\rightarrow 0} \nonumber
\end{eqnarray}
Therefore, for the 2nd derivative to be discontinuous
 we need, 
$ \frac{4}{3} \leq \chi < \frac{3}{2} $. \\ \\
(iii) Similarly, for the $l^{th}$ derivative to be discontinuous, the necessary condition is $\frac {l+2}{l+1} \leq \chi < \frac{l+1}{l}$ \\ \\
(iv) Careful observation of Eq. (37)
reveals $\Delta ^{l}$ is independent of $j$ for $\chi=1$, which indicates $\Delta = 0$ for $\eta \rightarrow 0$. Thus if $\chi=1$ all derivatives of $C_{v}$ are continuous. Using all these information, one can find out the hierarchy of the condensation transition (see table 1).\\\\
\section{Appendix}
\begin{figure}[h!]
\centering
\includegraphics[ height=8cm, width=8cm]{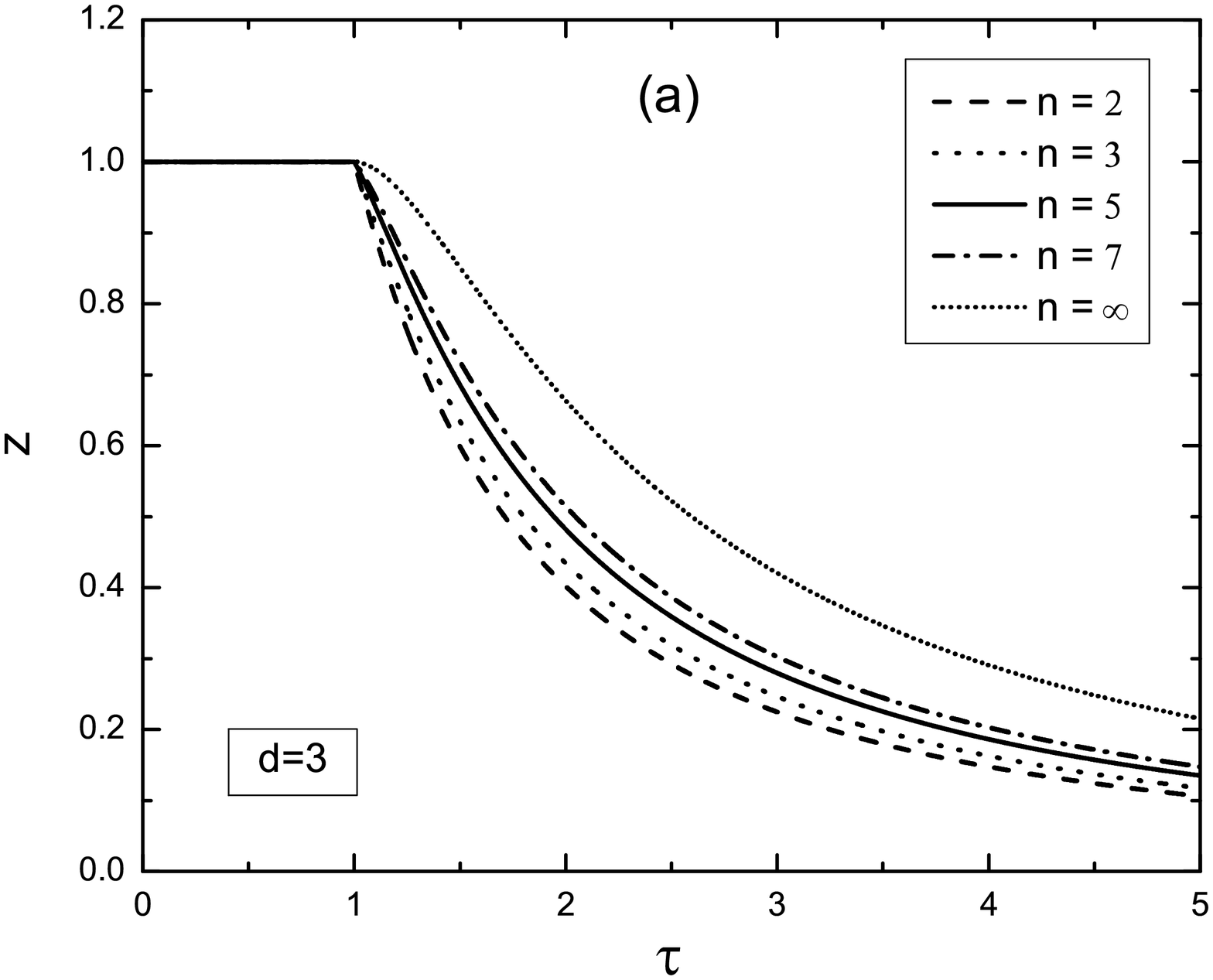}  
\includegraphics[ height=8cm, width=8cm]{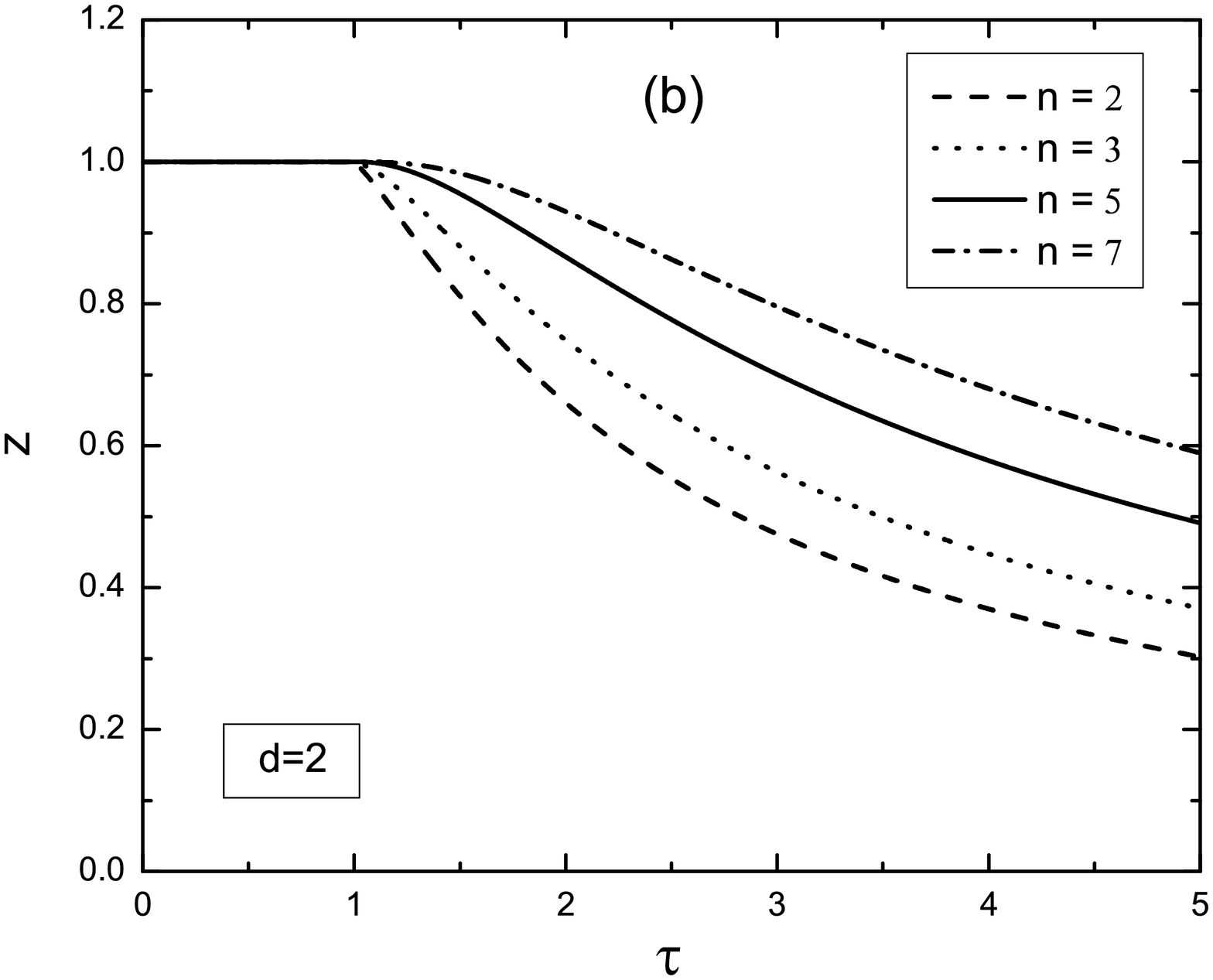}  
 \caption{Fugacity as a function of $\tau=\frac{T}{T_C}$, with different power law potentials. }
  \label{fig:boat1}
\end{figure}
In this section we demonstrate how fugacity can be expressed as a funtion of $\tau=\frac{T}{T_C}$ following
kelly's work.
The number of particles in the excited state near critical temperature,
\begin{eqnarray}
 N_e=\frac{1}{\lambda_c ^d}\zeta(\chi)\nonumber
\end{eqnarray}
And, the total number of particles can be written as,
\begin{eqnarray}
 N=\frac{1}{\lambda^d}g_\chi (z)\nonumber
\end{eqnarray}
In the noncondensed phase, one can approximate $N_e\simeq N$. So, in that case we get from the above two equations,
\begin{eqnarray}
 \frac{\tau^{d/2 }g_{\chi} (z)}{\zeta{(\chi)}}=1  \nonumber
\end{eqnarray}
Solving the above equation numerically, using mathematica we get our desired result (see figure 4)\\
Another very important relation used in deriving the different thermodynamic quantities are,
\begin{equation}
(\frac{\partial z}{\partial T})_{V}=-\chi\frac{z}{T}\frac{g_{x}(z)}{g_{x-1}(z)} 
\end{equation}
\section{Conclusion}
The changes in characteristics of Bose condensation of ideal Bose gas due to an external generic power law potential are studied from the grand potential. The presented theorems turn out to be very important for trapped Bose systems (non-relativistic). But it will be interesting to generalize these theorems for relativistic Bose gas. 

\section{Acknowledgement}
MMF would like to thank  Dr. Jens Roder
for his hospitality during the author's visit in ISOLDE, CERN where a major part of the work is done.
Also the authors would like to thank  Mr. Arya Chowdhury for his cordial help to present the work.

  \end{document}